# Thermal Weibel instability induced magnetic fields co-exist with linear wakes in laser-ionized plasmas

Yipeng Wu[1], Audrey Farrell[1], Mitchell Sinclair[1], Chaojie Zhang[1], Irina Petrushina[2], Navid Vafaei-Najafabadi[2], Marcus Babzien[3], William Li[3], Igor Pogorelsky[3], Mikhail Polyanskiy[3], Mikhail Fedurin[3], Karl Kusche[3], Mark Palmer[3], Ken Marsh[1] and Chan Joshi[1]

1 Electrical and Computer Engineering Department, University of California Los Angeles, Los Angeles, CA 90095, U.S.A.
2 Department of Physics and Astronomy, Stony Brook University, Stony Brookn, NY 11794, U.S.A.
3 Accelerator Test Facility, Brookhaven National Laboratory, Upton, NY 11973, U.S.A.

*Abstract*

When a moderately intense, few-picoseconds long laser pulse ionizes gas to produce an underdense plasma column, a linear relativistic plasma wave or wake can be excited by the self-modulation instability that may prove useful for multi-bunch acceleration of externally injected electrons or positrons to high energies in a short distance. At the same time, due to the anisotropic temperature distributions of the ionized plasma electrons, the Weibel instability can self-generate magnetic fields throughout such a plasma on a few picosecond timescale. In the present paper we first show using simulations that both these effects do indeed co-exist in space and time in the plasma. Using our simulations, we make preliminary estimates of the transverse emittance growth of an externally injected beam due to the Weibel magnetic fields. We then present results of an experiment that has allowed us to measure the spatiotemporal evolution of the magnetic fields using an ultrashort relativistic electron probe beam. Both the topology and the lifetime of the Weibel instability induced magnetic fields are in reasonable agreement with the simulations.

1. Introduction

A wake is a disturbance trailing behind a driver as it propagates through a fluid [1]. A high-intensity ($a_0>2$) short ($\tau< \pi c/\omega_p$) laser pulse or a high-density ($n_b > n_p$), high-energy ($\gamma \gg 1$) and tightly focused ($\sigma_r$ and $\sigma_z < c/\omega_p$) charged particle bunch can excite a relativistic wake ($v_{ph} \sim c$) in an ionized gas - in a process known as laser wakefield acceleration (LWFA) or plasma wakefield acceleration (PWFA), respectively [2-4]. Here $a_0$ is the normalized vector potential of the laser pulse, $\tau$ is the pulse duration, $n_p$ is the plasma density, $c/\omega_p$ is the plasma skin depth, $\gamma$ is the relativistic Lorentz factor corresponding to the phase velocity of the wake $v_{ph}$, $n_b$ is the peak density of the charged particle bunch, and $\sigma_r$ and $\sigma_z$ are the r.m.s bunch length and radius. Highly nonlinear (plasma density modulation $n_1/n_p \geq 1$), plasma wakes are considered for high-gradient electron acceleration [5-9], Betatron X-ray radiation generation [10, 11], delivering electrons for radiotherapy [12] and producing high-brightness beams for compact free electron lasers [13-15]. At the other extreme, linear wakes ($n_1/n_p \sim O(0.1)$) excited by a longer and modest-intensity laser pulse ($a_0$<1) or a medium-current ($I_{peak}$< few kA) electron bunch can excite a quasi-linear plasma wave via the self-modulation instability. Here the laser pulse or the electron bunch is typically 1-few picoseconds long and can be used to accelerate positrons or electrons injected in a single or multiple wave buckets. This is possible because such linear relativistic wakes can oscillate coherently in plasmas for tens of picoseconds, i.e., long after the driver has passed. Therefore, it behooves us to investigate what other process(es) occurring due to self-organization of the plasma itself may interfere with the acceleration of particles by the wake. An example of one such effect is the generation of long-wavelength magnetic field structures [16,17] that can grow and saturate in the plasma on this timescale. These magnetic fields could increase the emittance of an accelerated beam by deflecting particles via the $\boldsymbol{v} \times \boldsymbol{B}$ force. Although wakefield generation by the self-modulation instability has been studied experimentally by various diagnostic techniques [18-23], the self-generated magnetic fields in a finite radius (bounded by the slowly expanding sheath) plasma had not been hitherto observed mainly because a suitable experimental technique to measure these fields did not exist until recently [24].

In this paper we first use three-dimensional (3D) particle-in-cell (PIC) simulations to show that a moderately intense laser pulse ($a_0 \leq O(1)$) can ionize a gas to produce a plasma, generate a wake (via the self-modulation instability) and produce magnetic fields (via the so-called thermal Weibel instability [16, 17]). We then show the results of an experiment that have allowed us to unambiguously measure the spatiotemporal evolution of the magnetic fields in such a plasma generated by a picosecond $CO_2$ laser pulse with high spatiotemporal resolution by imaging the deflections of an external relativistic electron probe beam. These fields grow on a timescale of several picoseconds and can therefore have the potential to co-exist with the wake. We find that the topology, magnitude, and evolution of these fields are consistent with the thermal Weibel instability expected in a transversely bounded plasma. Wakefield generation is a nonlinear effect ($\propto \nabla E^2$) whereas the generation of magnetic fields in the plasma is an example of self-organization of plasma currents arising from kinetic effects [24]. Our PIC simulations show that these two physically different phenomena can spatially and temporally co-exist.

Here we consider the self-modulated LWFA [25-27] rather than the normal LWFA because the shortest pulse duration that currently can be achieved for a $CO_2$ laser (wavelength ~9.2 μm) is ~2 ps (FWHM). The reason for using a $CO_2$ laser is that for a relatively low-density plasma one can generate a relativistic wake necessary for acceleration of externally injected modest energy charged particles to high energies (tens to hundreds of MeV) in a relatively short (few mm) distance. For the simulation shown in this paper, the plasma density is $n_p$=1x10$^{17}$cm$^{-3}$ which gives wavelength of a relativistic wake to be $2\pi c/\omega_p$ = 105.5 um, far shorter than the ~750 um (FWHM) pulse width of the laser pulse. When the pulse length of the laser is many wake periods long, the self-modulation instability can both modulate the laser pulse at the plasma period and excite a plasma wake. We therefore first briefly discuss the physics behind the self-modulation instability and the thermal Weibel instability.

The self-modulation instability [25-27] is a multi-dimensional instability that reduces to the forward Raman scattering instability [28] in one-dimensional (1D) case when the laser spot size is larger than the plasma wake wavelength. The instability grows from either the plasma noise (e.g., collective Thomson scattering from longitudinal density fluctuations in the plasma) or intensity fluctuations in the laser pulse. In the case of the forward Raman scattering instability, both frequency up (anti-Stokes)- and down (Stokes)-shifted scattered photons travel in the same direction as the incident photons because of phase matching. Interference of the incident and the scattered photons reinforces the density fluctuations and thus scatters more light. This causes the plasma wave and the scattered light waves to grow exponentially in the forward direction as in a classic four-wave parametric instability [28]. In the case of the self-modulation instability, the noise fluctuations in the plasma density or on the laser pulse alternately focus and defocus the incident laser, causing amplitude modulation of the laser envelope to grow as the laser pulse propagates in the plasma. Both these instabilities produce Stokes and anti-Stokes sidebands to the laser frequency each separated from the next sideband by the plasma frequency. In the special case where the plasma is long compared with the laser pulse length, the forward Raman instability is a spatiotemporal instability - that is to say that the growth rate of the instability depends not only on laser pulse duration but also on how far the laser pulse has propagated into the plasma [29].

Depending on the plasma density the laser pulse can also undergo relativistic self-guiding if the laser power $P$ exceeds the critical power for self-focusing ($P > P_{cr}$) where $P_{cr}$ [GW] = $17.4 n_c/n_p$, and $n_c = 1.32 \times 10^{19} \text{cm}^{-3}$ is the critical plasma density for a 9.2 um-wavelength laser) [30]. In the paper by Esarey et al. [26], the authors asserted that relativistic self-focusing was critical to the self-modulation instability. They stated that $P/P_{cr}$ needed to be larger than roughly 1/2 (depending on the sharpness of the rise time) in order for strong self-modulation instability to occur. However in a later work [27] it was shown that for any given value of plasma density, a pulse length can be chosen such that the self-modulation (or forward Raman scattering) instability can still occur within a Rayleigh length $Z_r$. As we will see, the formation of a linear wake does not need the laser pulse to be relativistic as long as the lengths of the pulse and the plasma are sufficiently long so that the instability can grow from noise by several e-foldings.

There are many similarities between a wake generated by a single short intense laser pulse and that generated by self-modulation of a longer intense laser pulse. In both cases the wake is relativistic, and in the linear regime it has a longitudinal electric field that scales as $\sqrt{n_e}$ V/cm. This accelerating structure is suitable for accelerating both electrons and positrons. Electrons can be accelerated in the region where the electric field of the wake is negative whereas the positrons can be accelerated in the positive field region. However, the longitudinal field has a variation in the transverse direction which makes extraction of the energy from the wake (by beam loading) while keeping a very narrow energy spread very challenging. One option is to make the wake one dimensional, but then the energy extraction is rather inefficient. Another option is to shape the accelerated beam, which has proven extremely challenging to do. A third option is to split the electron or positron bunch into many microbunches that are separated by ~ $2\pi c/\omega_p$. This concept of acceleration of tens of microbunches contained within a single macrobunch has been demonstrated in a warm prototype of an S/X-band conventional accelerator for the Next Linear Collider (NLC) [31, 32]. In principle, a similar technique could be employed in a plasma accelerator to increase the total charge that can be accelerated and improve the wake-to-macrobunch energy extraction efficiency as long as no other mechanisms can generate large electric or magnetic fields in the plasma that would reduce the brightness of the accelerated electron bunchlets by increasing their transverse emittance.

So what process can lead to the spontaneous generation of large-scale ($O(c/\omega_p)$) electric or magnetic fields on a timescale (few $1/\omega_p$) that would interfere with the particles that are being accelerated in the multi bunch scenario described above? It turns out that if the rising edge of the drive laser pulse itself is used to produce the plasma, then kinetic effects--namely the Weibel and the current filamentation instabilities arising from the anisotropic/nonthermal electron velocity distribution (EVD) functions--can give rise to macroscopic magnetic fields through self-organization of electron currents that originate from the spatial noise present in the temperature anisotropy.

## 2. 3D PIC Simulations

Now we show a 3D PIC simulation using the code OSIRIS [33] that illustrates how both the self-modulation and the Weibel instabilities can evolve together in a plasma. This simulation mimics the experimental conditions of the experiment (described later in this paper) carried out at Brookhaven National Laboratory's Accelerator Test Facility (BNL-ATF). The plasma is formed by ionizing hydrogen gas emanating from a supersonic gas jet using a linearly polarized (45 degrees from the horizonal (x-z) plane) $CO_2$ laser pulse that propagates in the z direction. To see the evolutions of both the self-modulation instability generated wake and the thermal Weibel instability induced magnetic fields, the simulation used a large stationary window with dimensions of $2000 \times 480 \times 480 (c/\omega_0)^3 \approx 2928 \times 703 \times 703$ (μm)$^3$ in the z, x, and y directions, respectively, where $\omega_0 = 2\pi c/\lambda_0$ is the $CO_2$ laser frequency with $\lambda_0 = 9.2$ μm the wavelength. Atomic hydrogen gas was initialized inside the simulation box with atomic density linearly rising from zero at $z = 0$ to $n_p = 1 \times 10^{17}$ cm$^{-3}$ at $z = 1$ mm followed by a 1.6 mm-long density plateau, after which the density quickly drops to zero. In the other two orthogonal directions, the gas density is uniform. We note that in order to save the computation cost, the simulation box only covers the beginning 2.6 mm region of the ~6 mm long plasma (~4 mm long density plateau with ~1mm long density ramps on each side) in the actual experiment, which is also the region that is probed for the magnetic fields in the experiment. The $CO_2$ laser has a FWHM pulse duration of ~2 ps, a focal spot size of $W_0 \approx 100$μm, and an energy of ~0.2 J, leading to a peak intensity of ~6×10$^{14}$ W/cm$^2$ and a normalized vector potential of $a_0 \approx 0.2$. The vacuum focal plane of the $CO_2$ laser is set to z=3 mm. The laser pulse duration is sufficiently long for the laser to dissociate the hydrogen gas followed by ionization of the atomic hydrogen. Therefore, in the simulations, the ionization of atomic hydrogen is calculated using the Ammosov–Delone–Krainov (ADK) model [34]. The plasma electrons are sampled by 16 macroparticles per cell and the ions are assumed to be immobile.

### 2.1 Anisotropic Plasma Temperatures

The phase space evolution of the ionized electrons 2.8ps, and 35.8ps after the passage of the peak of the laser pulse in the center of the simulation box (defined as time zero) is shown in Fig. 1a-b. The corresponding EVD evolution at both times is also plotted in Fig. 1a-b. By fitting a Gaussian function to the EVD, we can obtain the evolution of plasma temperatures in the three dimensions $T_{e\parallel}, T_{e\perp}$, and $T_{ez}$ (see Fig. 1c), where the subscripts || and ⊥ denote the directions parallel and perpendicular to the laser polarization direction in the x-y plane, respectively. As clearly shown, the plasma electrons at t=2.8 ps are hot in the laser polarization direction but cold in the other two orthogonal directions. Also given are the temperature anisotropy factors $A_{\parallel\perp} = T_{e\parallel}/T_{e\perp} - 1$ and $A_{\parallel z} = T_{e\parallel}/T_{ez} - 1$ (see Fig. 1d). At t=2.8ps the plasma is highly anisotropic with temperature anisotropy factors $A_{||\perp}$ and $A_{||z} > 10$. Thereafter $A_{||\perp}$ and $A_{||z}$ keep dropping due to the isotropization of the plasma driven by kinetic instabilities [35].

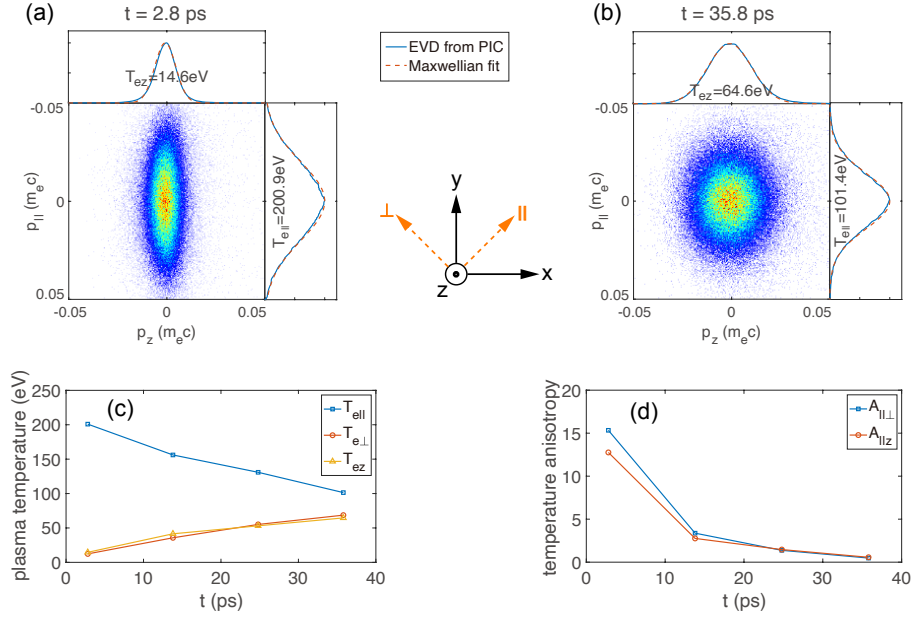

Fig. 1 (a) and (b) show the phase space as well as the EVD distributions of the plasma electrons from $z$=1.85mm to $z$=2.15mm at $t$=2.8ps and $t$=35.8ps, respectively. The evolutions of the temperature and anisotropy of the plasma electrons are shown in (c) and (d), respectively.

## 2.2 Evolution of the Plasma Wake

Figures 2a and 2b show the plasma density $n_e$ and the longitudinal electric field $E_z$ at different times, respectively. Clearly, $n_e$ shows plasma bounded by a sheath that separates it from the surrounding neutral gas. The width (transverse size) of the plasma is ~300 um, which is relatively narrow (just a few plasma wavelengths). The sheaths have a large transverse electric field ($E_y$ ~ 100 MV/m) but extremely localized, and electrons that have energies less than the sheath potential are reflected by the sheath electric field. Also one can see from Fig. 2b that there is a quasi-sinusoidal plasma wake as a result of the self-modulation instability of the $CO_2$ laser pulse. At the earliest time shown here ($t$=13.5 ps) the laser pulse has already left the box. The wavelength of this wake, ~105.5 um, is equal to that of a relativistic plasma wave $2\pi c/\omega_p$ indicating that we are exciting a linear wake (see the black dashed lines in Fig. 2b). Even at $t$=35.8 ps the wavelength remains roughly the same, indicating that there is little phase mixing of the electron trajectories and the wake remains quasi-sinusoidal. Although the amplitude of $E_z$ later decays, it's initial value of ~200 MV/m agrees with the theoretical prediction of linear wake ($\sim a_0^2 E_p/2$, where $E_p$ is the wave breaking limit of the cold plasma).

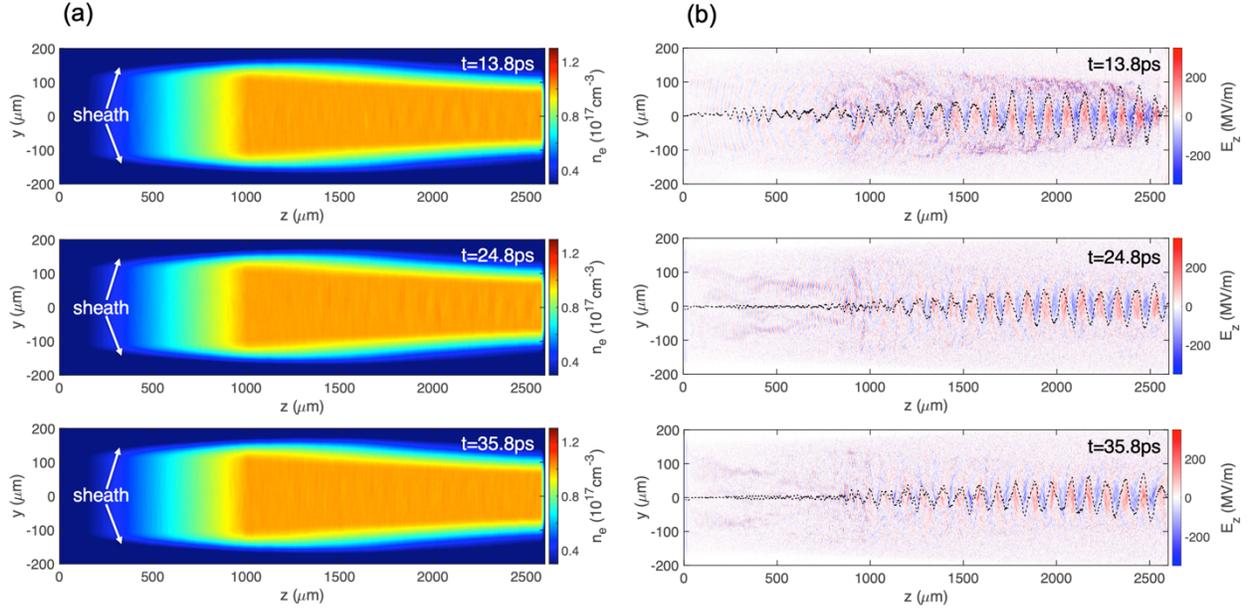

Fig. 2 The plasma density $n_e$ (a) and the longitudinal electric field $E_z$ (b) in the *y-z* plane (*x*=0) at *t*=13.8ps (top row), *t*=24.8ps (middle row), and *t*=35.8ps (bottom row), respectively. The laser pulse propagates from the left to the right. The black dashed line in each frame of (b) shows the corresponding smoothed central lineout of $E_z$ at *y*=0. Note that the density modulation is weak in (a), but the plasma wake can be seen more clearly in the electric field plots shown in (b). The short-wavelength wake feature in the upramp is expected from the wake recurrence phenomenon observed in a density upramp in Ref. [36].

## 2.3 Evolution of the Weibel Magnetic Fields

In addition to the electric fields of the linear plasma wake driven by the self-modulation instability, magnetic fields are expected to be simultaneously generated via the thermal Weibel instability. We characterize the instability as being thermal because it arises from the temperature anisotropy (temperatures differ in different spatial directions) as originally envisioned by Weibel [16]. Despite being one of the earliest kinetic instabilities to have been anticipated, it had not been conclusively observed in a laboratory plasma because of the difficulty of initializing the plasma with a known temperature anisotropy as postulated by Weibel. A major breakthrough occurred when high-field ionization was shown to generate nearly 100% ionization of a particular bound electron in just a few laser cycles. Such plasmas were shown to be extremely nonthermal and anisotropic. Furthermore, the degree of anisotropy could be controlled by controlling the laser wavelength, polarization, charge state etc [24]. The instability isotropized the plasma electrons and in the process a small fraction of the energy was converted to magnetic fields. It should be mentioned that collisions play little or no role in the thermalization. For instance at a relatively rarified density ($1 \times 10^{17}$ cm$^{-3}$) and a temperature of ~100 eV of the plasma shown in Fig. 2, the collision time [37] (O(~ns)) is much longer than 35.8 ps, the time at which the last frame in Fig. 2b is shown.

Figure 3a shows the transverse magnetic fields $B_y$ at different times. As one can see, $B_y$ is transversely bounded by the plasma sheath, and in all frames it has a quasi-1D periodic distribution along the $z$ direction, although it is initially somewhat irregular. This means that a quasi-single mode of the magnetic field has formed very quickly, and such a mode can maintain its topology for a long period. This is different from the evolution of Weibel magnetic fields in a wide (plasma width much larger than the plasma wavelength) plasma, where the initial morphology of the magnetic field is a 2D fish-net structure [38]. In a wide anisotropic plasma, the counter-propagating microscopic currents repel each other whereas the co-propagating microscopic currents attract each other. This coalescence of the currents keeps amplifying the magnetic field. Eventually current filaments are formed primarily in the cold-temperature directions (two orthogonal directions ($\perp$ and $z$) perpendicular to the laser polarization direction) and the magnetic fields also have wave vectors distributed in those directions, forming a 2D fish-net structure. However, in a bounded plasma, since the plasma is limited in size in the transverse ($\perp$) direction, instability modes with wavevectors along this direction and wavelengths larger than the plasma width cannot grow. Therefore, the most unstable growing mode is one with a wavevector in the longitudinal ($z$) direction, resulting in a quasi-1D periodic distribution of magnetic fields. We also note that the plasma currents keep merging as the instability grows and thus the spacing between the current filaments slightly increases, leading to an increase in wavelength of the magnetic fields as a function of time (see Fig. 3a). This is confirmed by the $k$-spectra obtained by Fourier transforming the magnetic fields $B_y$ in space, where $|k_z|$ shifts toward a smaller value (see Figs. 3b and 3c). In addition, as shown in Fig. 3d, the amplitude of $B_y$ first increases as the instability grows and then reaches saturation to ~0.5 Tesla. After that, $B_y$ decays with time.

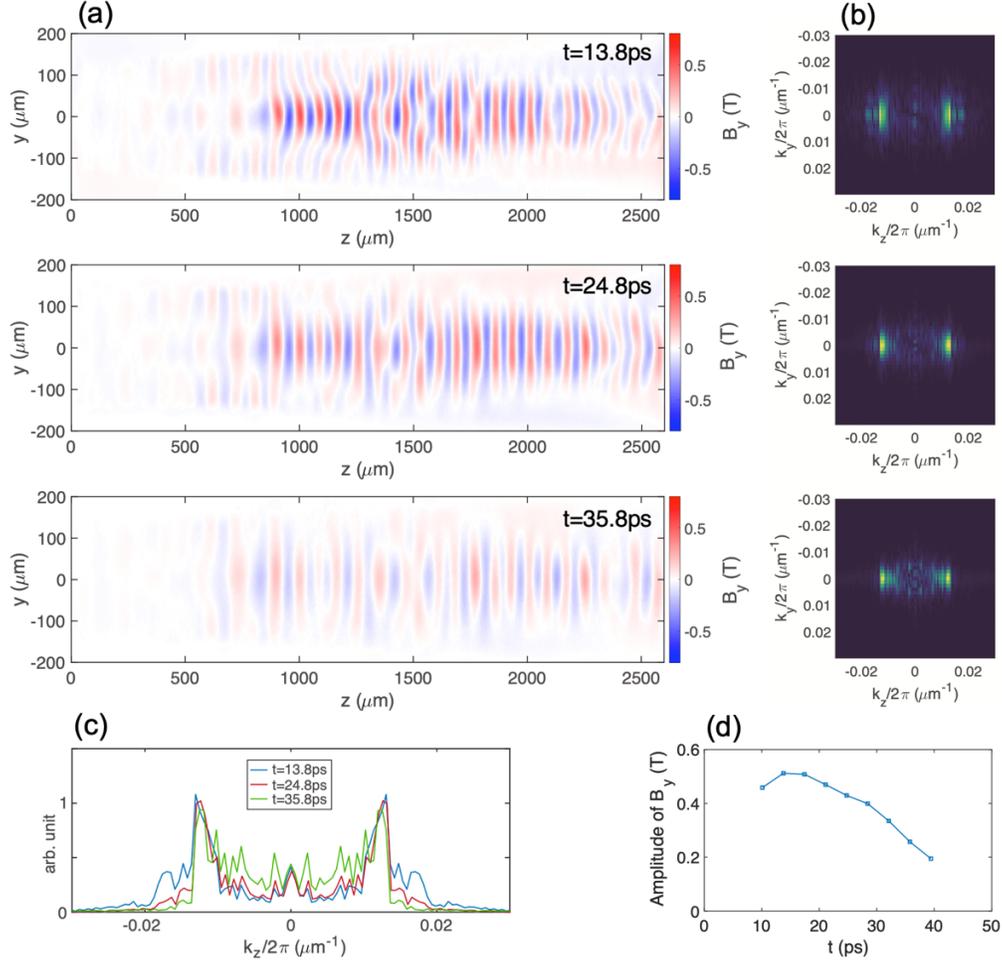

Fig. 3 The transverse magnetic field $B_y$ (a) in the y-z plane (x=0) and the corresponding 2D k-spectrum (b) obtained by taking the 2D Fourier transform of $B_y$ at t=13.8ps (top row), t=24.8ps (middle row), and t=35.8ps (bottom row), respectively. (c) The integrated $k_z$-spectrum corresponding to (b). (d) The evolution of the amplitude of $B_y$.

In the simulations, we have probed the thermal Weibel magnetic fields and created synthetic images by mimicking the electron radiography technique that we employ in the experiment. The probing is done by sending an ultrashort (1ps) relativistic (50.5MeV) electron beam through the plasma at various times and recording the deflections of the electrons by the magnetic fields. In our case if an electron probe with an initially homogeneous charge distribution is sent through the plasma along the +x direction, the electrons will be mainly deflected by the magnetic field components $\boldsymbol{B_y}$ and $\boldsymbol{B_z}$. The $\boldsymbol{B_y}$ field will cause deflections in the z direction and the $\boldsymbol{B_z}$ field will produce deflections in the y direction. These deflections then form density structures after propagating in vacuum for an optimum distance. This probing configuration is insensitive to deflections by the oscillating electric fields of the plasma wake because these fields oscillate at the plasma frequency $\boldsymbol{\omega_p}$, and as the picosecond-long probe travels through the plasma, the contributions of the wake are averaged over many periods ($\omega_p^{-1} = 53 \text{ fs}$) thus approaching to zero [39]. Figure 4 shows the modulated density profiles of the electron beam at a plane located 6 mm after the plasma at different time delays, from which we can obtain the evolution of the topology and strength (amplitude) of the magnetic fields. The vertical density strips are consistent with the quasi-1D periodic structure of the magnetic field $B_y$ along the $z$ direction (see Fig. 4). The spacing between the density strips slightly increases with time, indicating a reduction in the wavevector ($k_z$) value of the magnetic field. The modulation magnitude of the density strips also increases with time and saturates at $t \approx 23.8$ ps, after which it begins to decrease. We note that the density distribution of the electron probe, apart from exhibiting a vertical stripe pattern along the $z$ direction, is not symmetric in the $y$ direction. The region with negative $y$-values has a higher electron density compared to the region with positive $y$-values (as indicated by the red box in Fig. 4). We find that this is mainly caused by the longitudinal component $B_z$ of the Weibel magnetic fields.

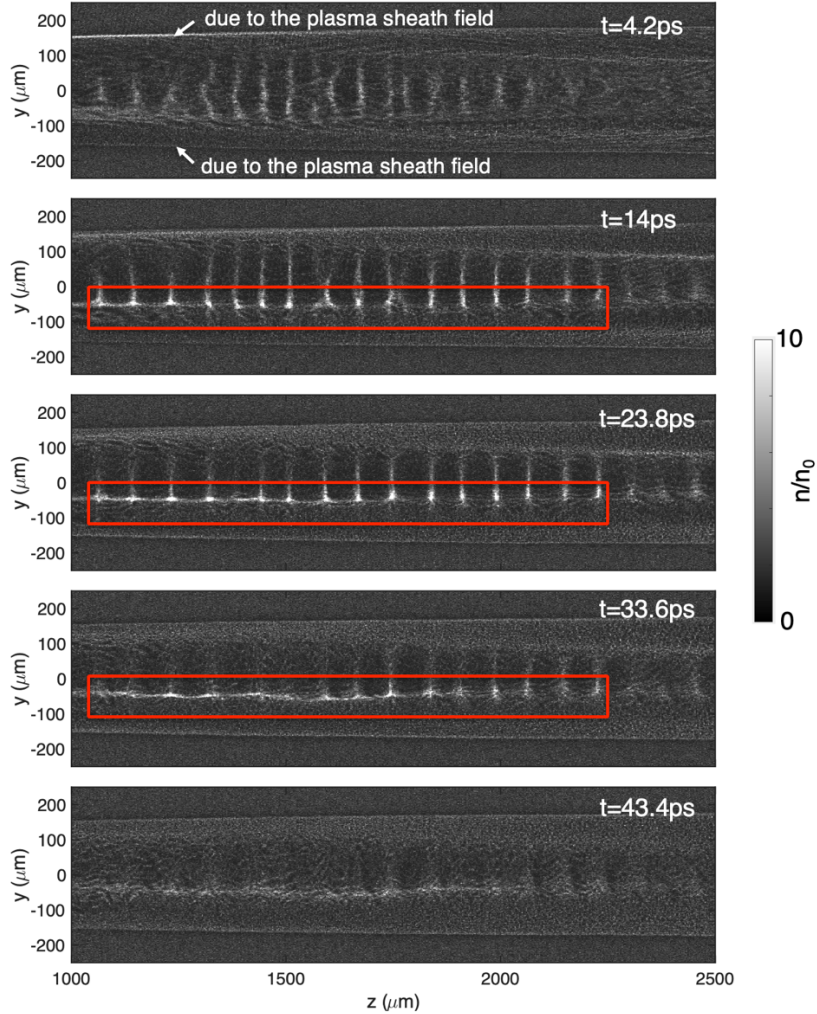

Fig. 4 Synthetic probe image at five different time delays. Note that in the synthetic image the electron beam is modulated by the transverse electric field $E_y$ of the plasma sheaths and by both the transverse ($B_y$) and longitudinal ($B_z$) components of the Weibel magnetic fields. The Weibel

magnetic fields are contained within the plasma sheath field. The red box indicates that the region with negative *y*-values has a higher electron density compared to the region with positive *y*-values.

## 3. Experiments

We have carried out experiments at BNL-ATF to measure the magnetic fields induced by the Weibel instability in a laser-plasma configuration that is also expected to generate a linear wake via the self-modulation instability as discussed in Section 2. The experimental layout is shown in Fig. 5. Plasmas with anisotropic temperatures are generated by ionizing hydrogen gas using a 2-ps, linearly polarized (45 degree), high-energy (~1 J) CO2 laser pulses [40]. Ultrashort (~1 ps) relativistic (50.5 MeV) electron bunches produced by the ATF linear accelerator are used to probe the magnetic fields. The detailed parameters of the $CO_2$ laser, plasma and electron probe beam are approximately equal to those used in the above simulations. To achieve a high spatial resolution, a set of permanent magnet quadrupoles (PMQs [41, 42]) are utilized to relay and magnify the electron probe to a YAG:Ce scintillator screen, which can convert the modulated electron flux to an optical image. The PMQ set consists of four pieces (two identical pairs) of Halbach-type high-gradient quadrupoles. These two pairs of quadrupoles were individually mounted, enabling the adjustment of their separation to modify magnification and shift the object plane location (see Fig. 5). Both assemblies were mounted on a third motorized translation stage, providing the capability to move the PMQs away from the path of the electron beam. When the movable PMQs are inserted, the object plane is positioned approximately 50 ± 0.5 mm downstream from the plasma and the image is magnified by a factor of 3.7 (further magnified by the optical system to an overall magnification of ~17.8) with a resolution of 3.3 μm.

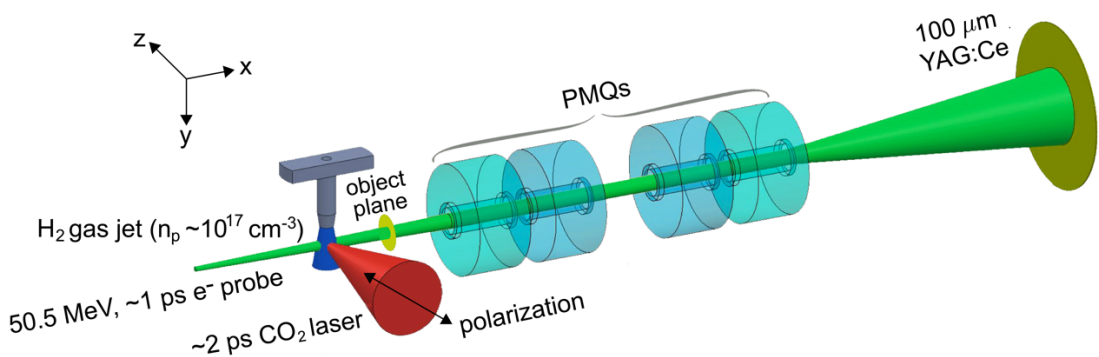

Fig. 5 Sketch of the experimental setup.

We note that we also used a ~70 fs (FWHM) 800 nm laser pulse to obtain shadowgraphs of the plasma wake. However, this proved challenging due to the plasma-induced refraction of the laser probe beam. By increasing the plasma density, Stokes and anti-Stokes sidebands in the $CO_2$ laser's spectrum became directly observable in the exact forward direction. In addition, evidence of the plasma wake emerged as sidebands at the plasma frequency in the collective Thomson scattering signal of the probe laser. Nevertheless, as the plasma density increased, the laser power approached the relativistic self-focusing threshold, leading to the rapid formation of a channel as plasma electrons were expelled. But there are other experiments conducted at powers below the relativistic self-focusing threshold and at intensities low enough that show long lived linear plasma wakes excited by single sub-100 fs, 800 nm laser pulse [36,43,44]. Therefore, for the rest of this paper, we will focus on the measurements of the Weibel magnetic fields.

### 3.1 Topology and Temporal Evolution of the Weibel Magnetic Fields

In the experiment, by adjusting the delay between the electron probe and the $CO_2$ laser, we have obtained a series of images of the density modulated probe beam due to deflections by the magnetic fields in the plasma. The corresponding raw images at different time delays after plasma formation are shown in Fig. 6a. As one can see, all the frames of electron density distribution have two horizontal (parallel to the laser propagation direction) boundaries, which are caused by the transverse sheath electric fields $E_y$ of the plasma. Between these two boundaries are the vertical density strips and their modulation amplitudes evolves as a function of time, indicating the evolution of the magnetic field amplitudes.

Based on the measured density distribution of the electron probe, we can retrieve the path integrals of the magnetic fields along the probe propagation direction. We first computationally retrieve the deflection angles of the probe electrons by solving an equivalent optimal transport problem (see Refs. [45, 46] for details). This method divides the beam profile into small regions called cells, and assumes that the deflection angles should be distributed in such a way that when mapping each cell of the source profile (the unperturbed electron beam profile) to the measured target profile (the beam profile at the object plane), the total displacement of all the cells should be minimized. Then, by assuming that the plasma has a slab geometry with a thickness of 300 μm, we can use the deflection angles to calculate the transverse magnetic fields. Fig. 6b shows the retrieved path-integrated magnetic fields $B_y$ at different time delays and Fig. 6c shows the corresponding 2D $k$-spectrum of $B_y$. The morphology of the magnetic fields in all frames are quasi-1D periodic distributions and the corresponding wavelength increases with time, which is consistent with the continuous decrease of $|k_z|$. These results are in good agreement with the simulation results shown in Fig.3. On the contrary, the measured magnitude of the magnetic field $B_y$ continuously increases with time to ~0.5 Tesla at $t$=50 ps, which is different from the simulation prediction that it first grows to the saturation value and then begins to decay. We believe that the reason is computational. In order to limit the computation time of our simulation, the number of macro-particles per cell used is relatively small (16 macro-particles per cell), resulting in certain artificial numerical collisions that cause the magnetic field to evolve faster than it should. Since the probing time window in the experiment is short compared with the evolution period of the instability, it does not capture the decaying process of the magnetic field.

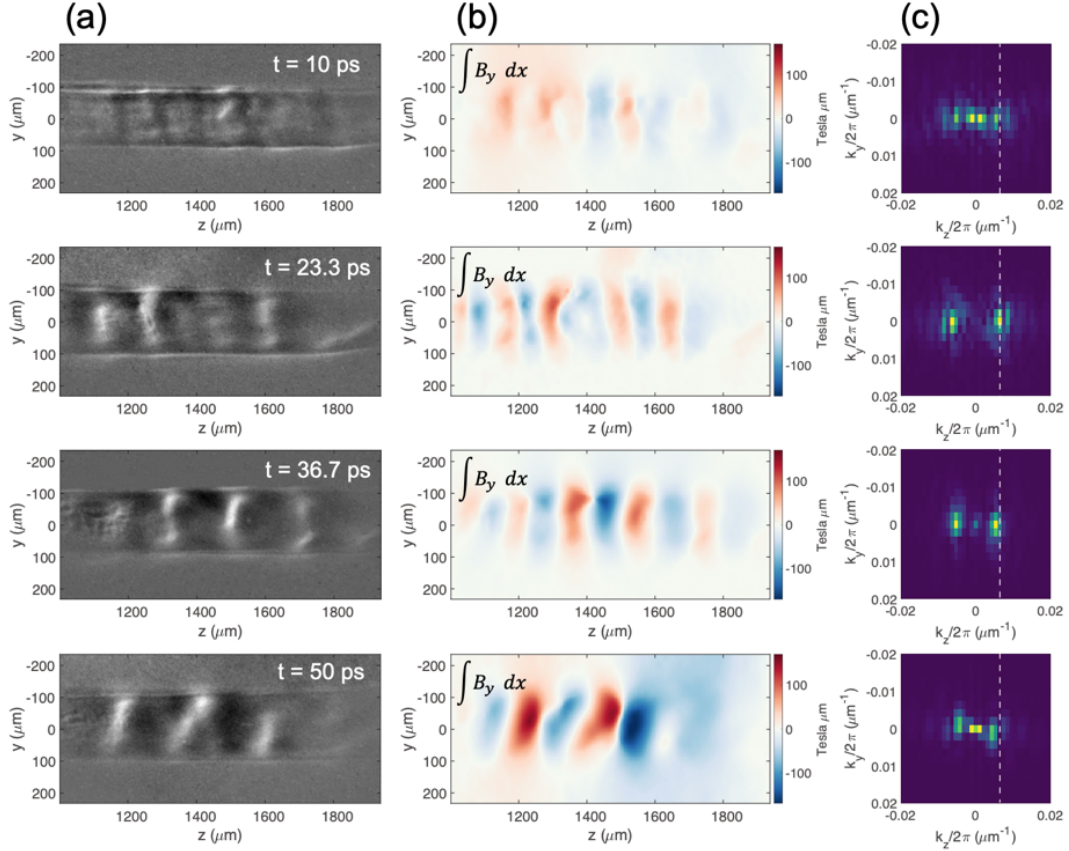

Fig. 6. Experimental results. The left column (a) shows the density modulations of the electron probe after being deflected by the magnetic fields in the plasma at different time delays. The middle (b) and right (c) columns show the corresponding retrieved $\int B_y\, dx$ and 2D $k$-spectrum of $B_y$, respectively.

## 4. Effect of Weibel Magnetic Fields on Accelerated Electrons

When both the linear plasma wakefield and the transverse Weibel magnetic field exist at the same time, the Weibel magnetic field will cause the deflection of the accelerated electron beam, which may lead to the deterioration of beam quality by increasing the transverse beam emittance. This effect will be the largest at low injection energies, such as at the very first stage of acceleration where a carefully prepared, few-MeV microbunched beam is injected into the linear wake [43]. To examine this, we initialize a witness electron beam that follows the $CO_2$ laser in the simulation. The witness beam has an energy of 10 MeV and a bi-Gaussian density profile with $\sigma_r = 10$ μm and $\sigma_z = 150$ μm. Its initial normalized emittance is set to 0.2 mm mrad. The time delay between the $CO_2$ laser and the witness beam is 10 ps. The simulation results show that the beam emittance growth over the course of the simulation is <35%. This is because the transverse magnetic field due to thermal Weibel instability is quasi-periodically distributed along the beam propagation direction like a magnetic undulator. This field undulates the electrons with both positive and negative values, which average out the transverse beam deflections. However, we note that, our simulation box only covers the beginning region of the experimental plasma. The emittance growth will be larger if the witness beam propagates for a longer distance in the plasma. The full impact of these fields on accelerated beam quality requires further careful investigation.

## 5. Summary

In summary, we have shown through self-consistent 3D PIC simulations that the thermal Weibel instability induced magnetic fields can co-exist with linear wakes in laser-ionized plasmas. We have measured the topology and the $k$-spectrum of the Weibel magnetic fields in experiment using an ultrashort relativistic electron probe. The retrieved spatiotemporal evolution of the magnetic field is in good agreement with the simulation results. Our preliminary results also show that the overall deterioration of the accelerated beam quality due to thermal Weibel magnetic fields may be tolerable because the Weibel field acts as a short-wavelength quasi-periodic undulating magnetic field.

**Acknowledgements**

We acknowledge the interest of Professor Warren B. Mori of UCLA in our work on kinetic instabilities in photoionized plasmas. The authors would like to acknowledge the OSIRIS Consortium, consisting of UCLA and IST (Lisbon, Portugal) for providing access to the OSIRIS 4 framework. This work was supported by DOE grant DE-SC0010064:0011 and NSF grant 2003354:003 at UCLA.